\documentclass{elsart}
\usepackage{graphicx}
\def\textbf#1{{\bf #1}}
\def\textit#1{{\it #1}}

\begin{document}

% ------------------------------
\begin{frontmatter}

\title{Inflation and deflation in financial markets}
\author[Tokyo1]{Taisei Kaizoji\thanksref{contact}}
\address[Tokyo1]{%
 Division of Social Sciences, International Christian University, Osawa, Mitaka, Tokyo 181-8585, Japan}
\thanks[contact]{%
Corresponding author. E-mail: kaizoji@icu.ac.jp; \\
URL: http://members.jcom.home.ne.jp/ephys/e-index.html
}

\begin{abstract}
The aim of this paper is to show new empirical results on the statistical properties of absolute log returns, defined as the absolute value of the log return, in a stock market. We used the daily data of the Nikkei 225 index of the 28-year period from January of 1975 to December of 2002, and compared the statistical properties of the return and absolute log returns in the inflationary (bubble) period with those in the deflationary (anti-bubble) period. Our results show that the distribution of absolute log returns is approximated by the $q$-exponential distribution where $ q = 1.14 $, that is, a {\it power law} distribution, in the inflationary period from January of 1975 to December of 1989, and it is accurately described by the $q$-exponential distribution where $ q = 1 $, that is, an {\it exponential} distribution, in the deflationary period from January of 1990 to December of 2002. 
\end{abstract}
\end{frontmatter}
%\newpage 

\section{Introduction}
Financial markets have been receiving increasing attention from physicists [1-10]. One of the reasons for this is that recent empirical studies on financial time series have found power laws, which emerge from complex systems, for many of these series [1-4]. 
In the present study, we focus attention on inflationary bubbles which ended in crashes in financial markets. In the last two decades, the financial markets have been distinguished by increasingly frequent financial crashes. Unfortunately, although it is extremely important to elucidate the mechanisms of financial crashes, little is yet known [10]. 
We here present the results of an empirical analysis performed by taking a new approach to bubbles and their collapses. We conducted a quantitative investigation on the statistical properties of price fluctuations in the Nikkei 225 index, which is the arithmetic average of the stock prices of 225 large companies listed in the Tokyo Stock Exchange. Figure 1(a) shows the daily series of the Nikkei 225 index during the period from January of 1975 to December of 2002. The Nikkei 225 index reached a high of almost 40,000 yen on the last trading day of the decade of the 1980s, but then declined approximately 14,309 during the period from the first trading day of 1990 to mid-August of 1992, a drop of about 63 percent. Thus, we divide the time series of the Nikkei 225 index into two parts: the period of inflation (bubbles), from January of 1975 to December of 1989, and the period of deflation (anti-bubbles), form January of 1990 to December of 2002. We show that the price fluctuations in the inflationary period can be approximated by asymptotic {\it power laws} provided by non-extensive statistical mechanics, while the price fluctuations in the deflationary period obey {\it exponential laws} provided by Boltzmann-Gibbs statistical mechanics. 

\section{Empirical analysis} 

\subsection{Log returns}
We first investigated the basic statistics of the price fluctuations. We define price fluctuations $ R_t $ as the {\it log return}, $ R_t = \ln S_t - \ln S_{t-1} $, where $ S_t $ denotes the Nikkei 225 index on the date $t$. Figure 1(b) shows the daily returns of the Nikkei 225 index in the period from January 4, 1975 to December 30, 2002. Table 1 shows differences in the statistical properties between the inflationary period (January 4, 1975 to December 29, 1989) and the deflationary period (January 4, 1990 to December 30, 2002). The mean of the returns $ \langle R \rangle $ is positive in the inflationary period and negative in the deflationary period. The standard deviation of the returns $ \sigma $ in the deflationary period is about two times as high as that in the inflationary period. Note that we omit the data of November 20, 1987, when the Nikkei 225 index had the greatest negative return, as an extreme value because extreme values would strongly affect the estimation of skewness $ s = E(R_t - \langle R \rangle)^3/\sigma^3 $ and kurtosis $ k = E(R_t - \langle R \rangle)^4 /\sigma^4 $. The return distribution is negatively skewed during the inflationary period, and positively skewed in the deflationary period. Finally, the kurtosis of the return distribution in the inflationary period is higher than that in the deflationary period, indicating that the return distribution has a {\it fat tail} in the inflationary period, compared to that in the deflationary period. That is, there is a higher probability for extreme values in the return distribution in the inflationary period than in the deflationary period. 
\bigskip
\begin{table}[htbp]
\begin{center}
\begin{tabular}{c|cc}
\hline
Statistics & \textit{Inflation (1975-1989)} & \textit{Deflation (1990-2002)} \\ \hline
Mean & 0.0006 & -0.0005 \\ 
S.D. & 0.007 & 0.015 \\ 
Skewness & -0.3 & 0.3 \\ 
Kurtosis & 9.5 & 6.1 \\ \hline
\end{tabular}
\caption[Statistics]{{\bf Statistics of the return}}
\end{center}
\end{table}

\subsection{Absolute log-returns}
For the purposes of the present study, we define volatility as the absolute value of the log return $ V_t = |R_t| $. Figure 1(c) shows the absolute log-return $ V_t $ for the Nikkei 225 index for the 28-year study period. Figures 2(a) and 2(b) show the semi-log plots of the complementary cumulative probability distributions of the absolute log-returns for the Nikkei 225 index for both the inflationary and deflationary periods. It follows from Figure 2(a) that the complementary cumulative probability distribution of the absolute log-returns for the period of deflation is accurately described by an exponential distribution, 
\begin{equation}
  P(V > x) \propto \exp(- \beta x), 
%(1)
\end{equation}
in the whole range of the absolute log-returns $ V_t $. We estimate the parameter $ \beta$ using the least square method, which gives $\beta = 1.45$ $ (0.0) $. It also follows from Figure 2(b) that the complementary cumulative probability distribution of the absolute log-returns for the period of inflation deviates from the exponential distribution as the absolute log-returns becomes greater. Figure 3(a) shows the log-log plot of the distribution of the absolute log-returns in the inflationary period. This plot shows an asymptotic power law where the value of the absolute log-return is large, but deviates from the power law where the value of the absolute log-return is small. One approach to incorporating this deviation from the power law is to consider the $q$-exponential function [11]. We find that the complementary cumulative probability distribution of the absolute log-returns of the inflationary period is well approximated by the following $q$-exponential distribution: 

\begin{equation}
 P(V > x) \propto [1 - (1 - q) \beta x]^{1/(1-q)}. 
%(2)
\end{equation}
We estimate the parameters, $ \beta$ and $ q $ using the nonlinear least square method, which gives $\beta = 2$ and $ q = 1.14 $. Note that for $ q > 1 $ the $q$-exponential distribution has a power-law tail, i.e., $ P(V > x) \sim x^{-\alpha} $ [4], and for $ q = 1 $ it becomes an exponential function since $ \lim_{q\to1} \exp_q(- \beta x) = \exp(- \beta x) $. Here, we introduce another possible way of analyzing data by using a generalized mono-log plot based on the inverse function of the $q$-exponential function, which is the $q$-logarithmic function given by $ \ln_q[P(V > x)] \equiv (x^{1-q} - 1)/(1 - q) $. This generalized function arises naturally in the framework of Tsallis statistics [11,12]. It is easy to verify that the plot of $ \ln q[r(x)] $ versus $x$ leads to a straight line if the data are well described by the $q$-exponential distribution (Figure 3(b)). 

\section{Concluding remarks}
Our results indicate that the absolute log-return is approximated by the $q$-exponential distribution where $ q = 1.14 $ in the inflationary period, while during the deflationary period, the distribution of the absolute log-returns is accurately described by an {\it exponential} distribution, that is, the $q$-exponential distribution where $ q \to 1 $. 
As has been demonstrated by C. Tsallis [11], Boltzmann-Gibbs statistical mechanics typically provides {\it exponential laws} for describing stationary states, while non-extensive statistical mechanics typically provides asymptotic {\it power laws}. 
The empirical findings presented here indicate that the price fluctuations in the inflationary (bubble) period can be described by a power law based on non-extensive statistical mechanics, while the price fluctuations in the deflationary (anti-bubble) period can be described by an exponential law based on Boltzmann-Gibbs statistical mechanics. 
The present study raises the interesting question, left for future research, of whether the empirical facts found here hold true in other stock markets such as the New York Stock Exchange. This study will be left for future work. Our empirical results suggest that the momentous structural change was cased and the $q$-value changed from greater than unity to unity at the beginning of 1990 when the speculative bubble collapsed in Japan's stock markets. However, to date, no model so far has successfully explained such changes in the statistical properties of price fluctuations. Hence the next step is to model such behavior in stock markets. 

\section{Acknowledgements}
The author would like to thank Prof. Thomas Lux for valuable comments and criticism, and Ms. Michiyo Kaizoji for her assistance in collecting the data. 
Financial support from the Japan Society for the Promotion of Science under Grant-in-Aid No. 06632 is gratefully acknowledged. All remaining errors, of course, are mine.

\newpage
% ------------------------------
% Figure Captions
% ------------------------------

\begin{figure}
\begin{center}
  \includegraphics[height=19cm,width=14cm]{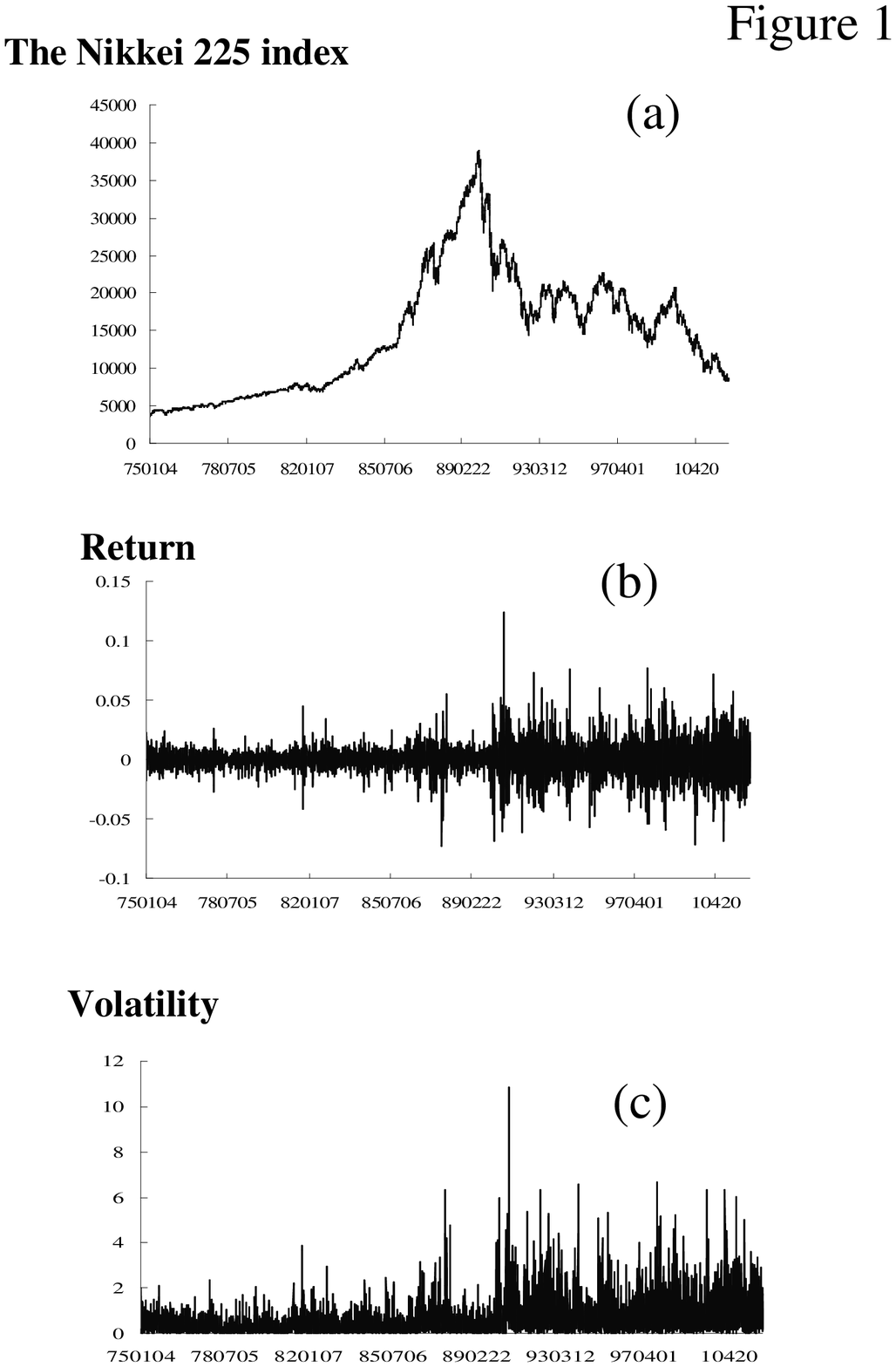}
\end{center}
\caption{(a) The time series of the Nikkei 225 index $ S_t $ in the period from January 4, 1975 to December 30, 2002. (b) The time series of the returns $ R_t = \ln S_t - \ln S_{t-1} $ for the Nikkei 225 index from January 4, 1975 to December 30, 2002. (c) The time series of the absolute log-returns $ V_t = |R_t| $ for the Nikkei 225 index in the period of January 4, 1975 to December 30, 2002. The period of inflation occurred from January 4, 1975 to December 29, 1989 and the period of deflation was from January 4, 1990 to December 30, 2002.} 
\label{fig1}
\end{figure}

\begin{figure}
\begin{center}
  \includegraphics[height=19cm,width=14cm]{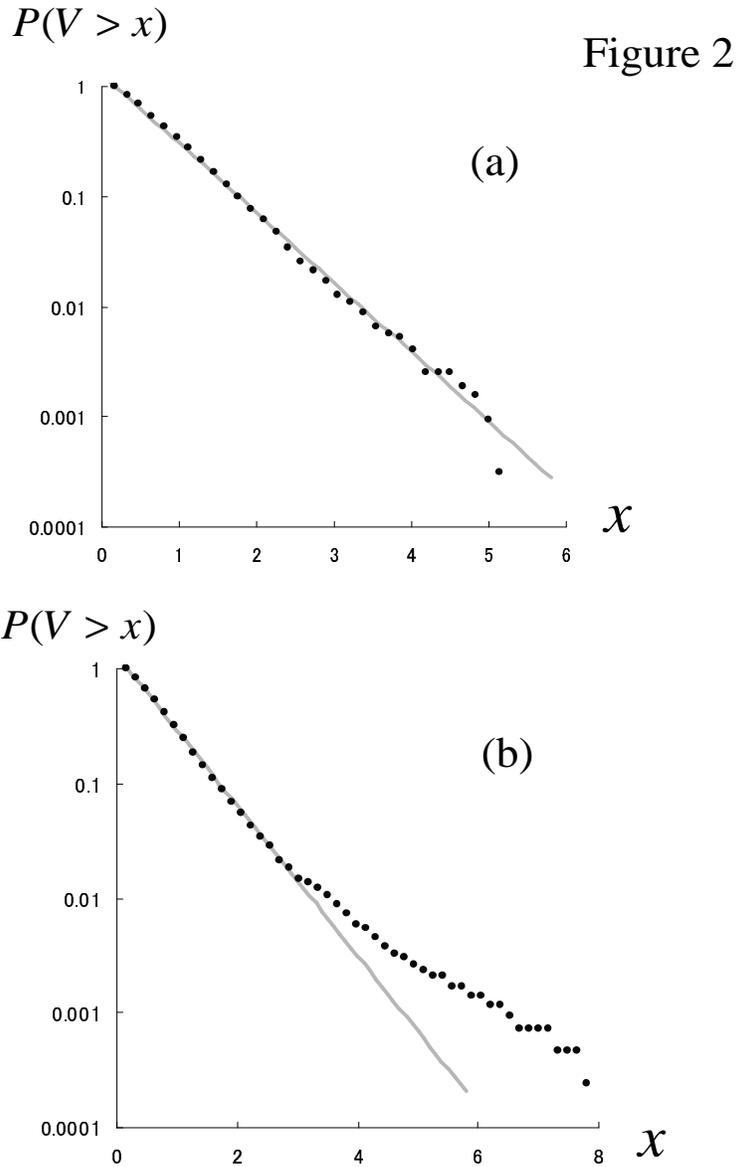}
\end{center}
\caption{(a) The semi-log plot of the complementary cumulative probability distribution of the absolute log-returns $ V_t $ for the Nikkei 225 index during the period of deflation (January 1990-December 2002). The straight lines indicate exponential functions $ P(V > x) \propto \exp(- \beta x) $ where $ \beta = 1.45 $. (b) The semi-log plot of the complementary cumulative probability distribution of the absolute log-returns $ V_t $ for the Nikkei 225 index in the period of inflation (January 1975-December 1989).}
\label{fig2}
\end{figure}

\begin{figure}
\begin{center}
  \includegraphics[height=19cm,width=14cm]{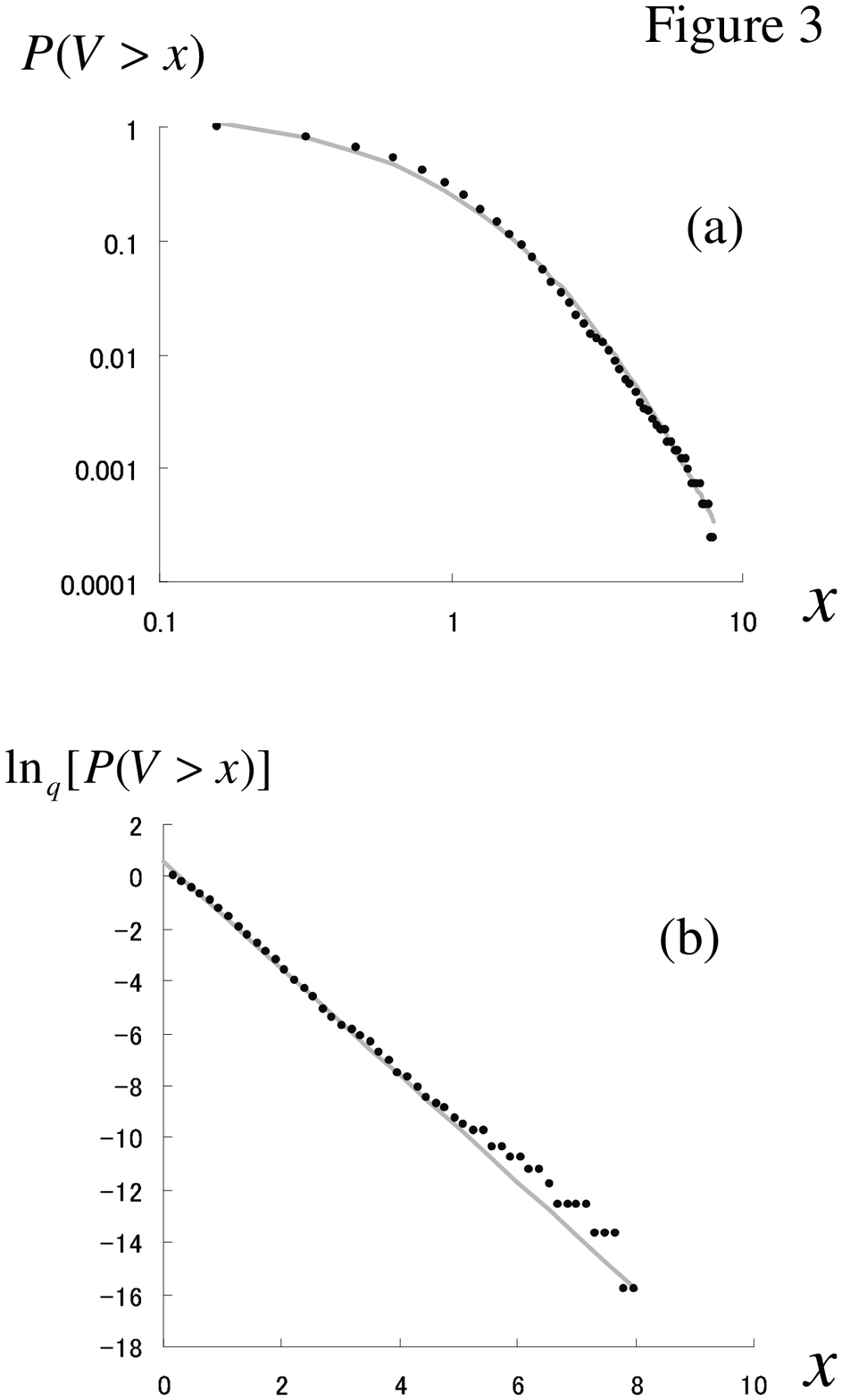}
\end{center}
\caption{(a) The log-log plot of the complementary cumulative probability distribution of the absolute log-returns $ V_t $ for the Nikkei 225 index during the period of inflation (January 1975-December 1989). The solid lines indicate $q$-exponential functions $ P(V > x) \propto [1 - (1 - q) \beta x]^{1/(1-q)} $ where $ \beta = 2 $ and $ q = 1.14 $. (b) The plot of $ \ln_q[P(V > x)] \equiv (x^{1-q} - 1)/(1 - q) $ versus $ x $. The coefficient of determination of the straight line on the generalized mono-log plot is equal to $ R^2 = 0.996 $.} 
\label{fig3}
\end{figure}

\end{document}